\documentclass{aip-cp}

\usepackage{etoolbox}
\makeatletter \patchcmd{\@@tablenote}{\xdef}{\protected@xdef}{}{}
\makeatother

\usepackage[numbers]{natbib}
\usepackage{rotating}
\usepackage{graphicx}
\usepackage{color}
\graphicspath{{art/}}

\begin{document}

\title{Bohm potential for the time dependent harmonic oscillator }

\author[aff1]{Francisco Soto-Eguibar}
\author[aff2]{Felipe A. Asenjo}
\author[aff3]{Sergio A. Hojman}
\author[aff1]{H\'ector M. Moya-Cessa}
\affil[aff1]{Instituto Nacional de
 Astrof\'{\i}sica, \'Optica y Electr\'onica, Calle Luis Enrique Erro
  No. 1, Santa Mar\'{\i}a Tonanzintla, Puebla, 72840, Mexico.}
\affil[aff2]{Facultad de Ingenier\'ia y Ciencias,
Universidad Adolfo Ib\'a\~nez, Santiago 7491169, Chile.}
\affil[aff3]{Departamento de Ciencias, Facultad de Artes Liberales,
Universidad Adolfo Ib\'a\~nez, Santiago 7491169, Chile.\\
Departamento de F\'{\i}sica, Facultad de Ciencias, Universidad de Chile,
Santiago 7800003, Chile.\\
Centro de Recursos Educativos Avanzados,
CREA, Santiago 7500018, Chile.}

\maketitle

\begin{abstract}
In the Madelung-Bohm approach to quantum mechanics, we consider a (time dependent) phase that depends quadratically on position and show that it leads to a Bohm potential that corresponds to a time dependent harmonic oscillator, provided the time dependent term in the phase obeys an Ermakov equation. 
\end{abstract}

\section{Introduction}
Harmonic oscillators are the building blocks in several branches of physics, from classical mechanics to quantum mechanical systems. In particular, for quantum mechanical systems,  wavefunctions have been reconstructed as is the case for quantized fields in cavities \cite{Bert}  and for ion-laser interactions \cite{Leib}. Extensions from single harmonic oscillators to time dependent harmonic oscillators may be found in shortcuts to adiabaticity \cite{Muga0}, quantized fields propagating in dielectric media \cite{Dodonov}, Casimir effect \cite{Ricardo} and ion-laser interactions \cite{Casanova2018}, where the time dependence is necessary in order to trap the ion.\\
Time dependent harmonic oscillators have been extensively studied and  several invariants have been obtained \cite{Lewis0,Lewis,Ray,Thylwe1998,JPA}. Also algebraic methods to obtain the evolution operator have been shown \cite{Cheng1988}. They have been solved under various scenarios such as time dependent mass \cite{Cheng1988,Moya2007,Ramos2018}, time dependent frequency \cite{Pedrosa,JPA}  and applications of invariant methods have been studied in different regimes \cite{Santos}. Such invariants may be used  to control quantum noise \cite{Muga} and to study the propagation of light in waveguide arrays \cite{Barral1,Barral2}. Harmonic oscillators may be used in more general systems such as waveguide arrays \cite{Leija1,Observation,Zarate}.\\
In this contribution, we use an operator approach to solve the one-dimensional Schrödinger equation in the Bohm-Madelung formalism of quantum mechanics. This formalism has been used to solve  the Schrödinger equation for different systems by taking the advantage of their non-vanishing Bohm potentials \cite{Hojman,Hojman2,Hojman3,makowski}. Along this work, we show that a time dependent harmonic oscillator may be obtained by choosing a position dependent quadratic time dependent phase and a Gaussian amplitude for the wavefunction. We solve the probability equation by using operator techniques. As an example we give a rational function of time for the time dependent frequency and show that the Bohm potential has different behavior for that functionality because an auxiliary function needed in the scheme, namely the functions that solves the Ermakov equation, presents two different solutions.

\section{One-dimensional Madelung-Bohm approach}
The main equation in quantum mechanics is the Schrodinger equation, that in one dimension and for a potential $V(x,t)$ is written as (for simplicity, we set $\hslash=1$)
\begin{equation}\label{SE}
i\frac{\partial \psi(x,t)}{\partial t}= -\frac{1}{2m}\frac{\partial^2 \psi(x,t)}{\partial x^2}+V(x,t)\psi(x,t)
\end{equation}
with $\psi(x,t)$ the wavefunction of the quantum mechanical system. We may give a solution in terms of a polar decomposition \cite{Wyatt,Holland,Hojman}
\begin{equation}\label{psi}
\psi(x,t)=A(x,t)e^{i S(x,t)}
\end{equation}
with $A(x,t)$ and $S(x,t)$ real functions that depend on time and position. We may separate the real and imaginary parts that come from substitution of (\ref{psi}) in (\ref{SE}), the first equation being a Quantum Hamilton–Jacobi equation (QHJE),  similar to its classical counterpart {(where the dot represents the time derivative and the prime the space derivatives)},
\begin{equation} \label{H-J}
\frac{1}{2m}S'^2+V_B+V+\dot{S}=0,
\end{equation}
and the second one, the continuity (probability conservation) equation,
\begin{equation}\label{Prob} 
\frac{1}{2m}(2A'S'+AS'')+\dot{A}=0,
\end{equation}
with the Bohm potential defined by \cite{Made,Bohm}
\begin{equation}\label{bohmpot}
V_b=-\frac{1}{2m}\frac{A''}{A}.
\end{equation}

\section{Operator approach to the solution of continuity equation}
The probability equation (\ref{Prob}) may be rewritten as a Schrodinger-like equation
\begin{equation}\label{Operator}
\frac{\partial A}{\partial t}=-\frac{1}{2m}\left(2S'\frac{\partial }{\partial x}+S''\right)A,
\end{equation}
that, by using the momentum operator $\hat{p}=-i\frac{\partial }{\partial x}$, we may write as
\begin{equation}\label{OperatorI}
\frac{\partial A}{\partial t}=-\frac{1}{2m}\left(i2S'\hat{p}+S''\right)A.
\end{equation}
Now, we choose {$S(x,t)=Q(x)\dot{\nu}(t)+\mu(t)$}, such that $S'=Q'\dot{\nu}$, and we use the property $[f(x),\hat{p}]=if'(x)$,
to rearrange the term
\begin{equation}
2S'\hat{p}=\dot{\nu}(Q'\hat{p}+Q'\hat{p})=\dot{\nu}[Q'\hat{p}+\hat{p}Q'+iQ''],
\end{equation}
so that we may write Eq. (\ref{OperatorI}) as
\begin{equation}
\frac{\partial A}{\partial t}=-i\frac{\dot{\nu}}{2m}\left(Q'\hat{p}+\hat{p}Q'\right)A.
\end{equation}
The above equation is readily solvable, with  solution
\begin{equation}
A(x,t)=\exp\left\lbrace -\frac{i}{2m}\int\dot{\nu}(t)dt 
\left[Q'(x)\hat{p}+\hat{p}Q'(x)\right] \right\rbrace A_0(x),
\end{equation}
where $A_0(x)=A(x,t=0)$, the initial condition, is an arbitrary (square integrable) function of position.\\
Next, we  assume $Q(x)=x^2/2$ and $m=1$ to find the solution
\begin{equation}\label{0110}
A(x,t)=\exp\left[-i\frac{\nu(t)}{2}\left( x\hat{p}+\hat{p}x\right) \right]A_0(x).
\end{equation}
In the above equation, the operator $\exp\left[ -i\frac{\nu(t)}{2}\left( x\hat{p}+\hat{p}x\right)\right] $
is the so-called squeeze operator \cite{Yuen,Caves}. By choosing an initial condition $A_0(x)=\pi^{-1/4} \exp\left(-x^2/2 \right)$,
\begin{equation}\label{0120}
A(x,t)=\frac{1}{\pi^{1/4}}\exp\left[-\frac{x^2}{2}e^{-2\nu(t)}-\frac{\nu(t)}{2}\right],
\end{equation}
where we have used the fact that $\exp\left[-i\frac{\nu(t)}{2}\left( x\hat{p}+\hat{p}x\right) \right]x\exp\left[i\frac{\nu(t)}{2}\left( x\hat{p}+\hat{p}x\right) \right]=x\exp[-\nu(t)]$ that is easily found from the Hadamard formula that states that for two operators $\hat{A}$ and $\hat{B}$, $e^{r\hat{A}}\hat{B}e^{-r\hat{A}}=\hat{B}+r[\hat{A},\hat{B}]+\frac{r^2}{2!}[\hat{A},[\hat{A},\hat{B}]]+\dots$.\\
Now we are ready to calculate the Bohm potential, for which we need $A'(x,t)=-xe^{-2\nu(t)}A(x,t)$ and $A''(x,t)=\left[{x^2e^{-4\nu(t)}}-{e^{-2\nu(t)}}\right]A(x,t)$,
to obtain
\begin{equation}\label{0130}
V_B(x,t)=-\frac{x^2}{2}\exp\left[-4\nu(t)\right]+\frac{1}{2}\exp\left[-2\nu(t)\right] .
\end{equation}
By choosing $\dot{\mu}=-\exp\left[-2\nu(t)\right] /2$, we obtain from  equation (\ref{H-J}),
\begin{equation}\label{0140}
V+\frac{x^2}{2}\left(\ddot{\nu}+\dot{\nu}^2-e^{-4\nu}\right)=0;
\end{equation}
by changing variables to $\rho=\exp\left(\nu\right) $, we get  $\dot{\mu}=-1/\left(2\rho^2\right)$, $\dot{\nu}=\dot{\rho}/\rho$ and
$\ddot{\nu}=\left( \rho\ddot{\rho}-\dot{\rho}^2\right)/2 $
that gives
\begin{equation}\label{0150}
V+\frac{x^2}{2\rho}\left(\ddot{\rho}-\frac{1}{\rho^3}\right)=0.
\end{equation}
If $\rho$ obeys the Ermakov equation
\begin{equation}\label{0160}
\ddot{\rho}+\Omega^2(t)\rho=\frac{1}{\rho^3},
\end{equation}
we end up with the potential for the time dependent harmonic oscillator
\begin{equation}
V(x,t)=\frac{\Omega^2(t)}{2}x^2.
\end{equation}
The wave function for  the above potential and conditions then reads
\begin{equation}
    \psi(x,t)=\frac{1}{\pi^{1/4}}\exp\left\{-\frac{x^2}{2}e^{-2\nu(t)}-\frac{\nu(t)}{2}+i\left[\frac{x^2}{2}\dot{\nu}(t)+\mu(t)\right]\right\}.
\end{equation}

\section{As example a rational function of time}
We consider a time dependent frequency of the form
\begin{equation}
\Omega(t)=\frac{1}{a+bt},
\end{equation}
that shows for which has different solutions for  $b=2$ as will be shown below.

\subsection{Case $0<b<2$}
\begin{figure}[htbp]
\centering
\includegraphics[width=0.9\linewidth]{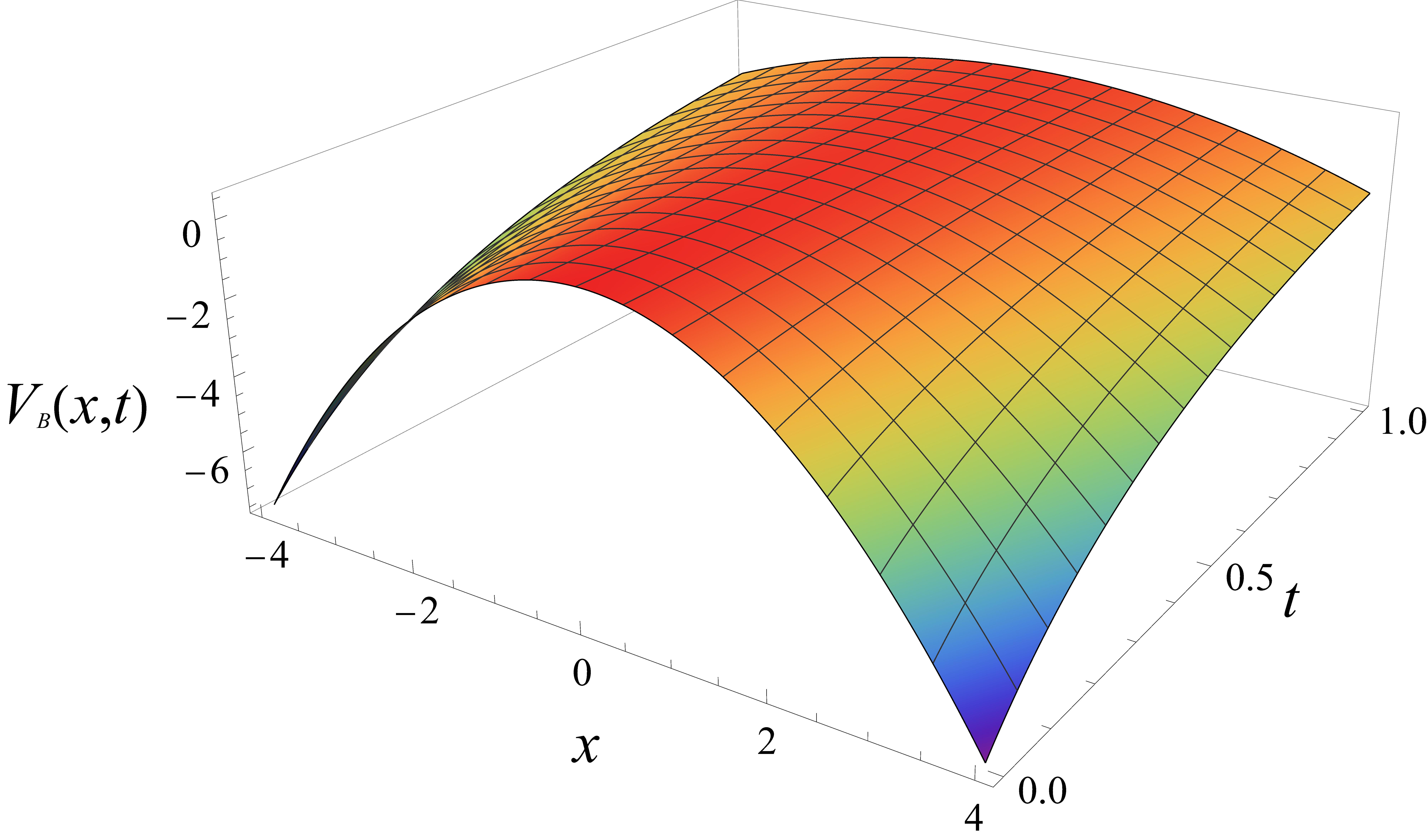}
\caption{The Bohm potential (\ref{0270}) for $b=1$.}
\label{fig1}
\end{figure}

The  solution for the Ermakov equation for the frequency chosen above gives the auxiliary function
\begin{equation}\label{0250}
\rho(t)=C\sqrt{a+bt},
\end{equation}
with $C=\left(1-b^2/4\right)^{-1/4}$. Eq. (\ref{0110}) sets the value for $\rho(0)=1$, which in turn forces $A=\sqrt{1-b^2/4}$ and gives the function
\begin{equation}\label{0260}
\nu(t)=-\frac{1}{4}\ln \left(1-\frac{b^2}{4}\right) +\frac{1}{2}\ln \left[\left(1-\frac{b^2}{4}\right)^{1/2}+Bt\right].
\end{equation}
The Bohm potential, Eq. (\ref{0130}), is then 
\begin{equation}\label{0270}
V_B(x,t)=-{\frac{1-b^2/4}{2(a+bt)^2}}x^2+\frac{\sqrt{1-b^2/4}}{2(a+bt)},
\end{equation}
that shows that there exists a Bohm potential only for $0<b<2$.\\
For $0<b<2$ and $t\geq0$, the Bohm potential (\ref{0270}) is well defined, without singularities, and has the form shown in Fig. \ref{fig1}. The Bohm potential exhibits the same qualitative behavior for all values of the parameter $b$ in the interval $(0,2)$.
\\

\subsection{Case $b=2$}
\begin{figure}[htbp]
\centering
{\includegraphics[width=0.9\textwidth]{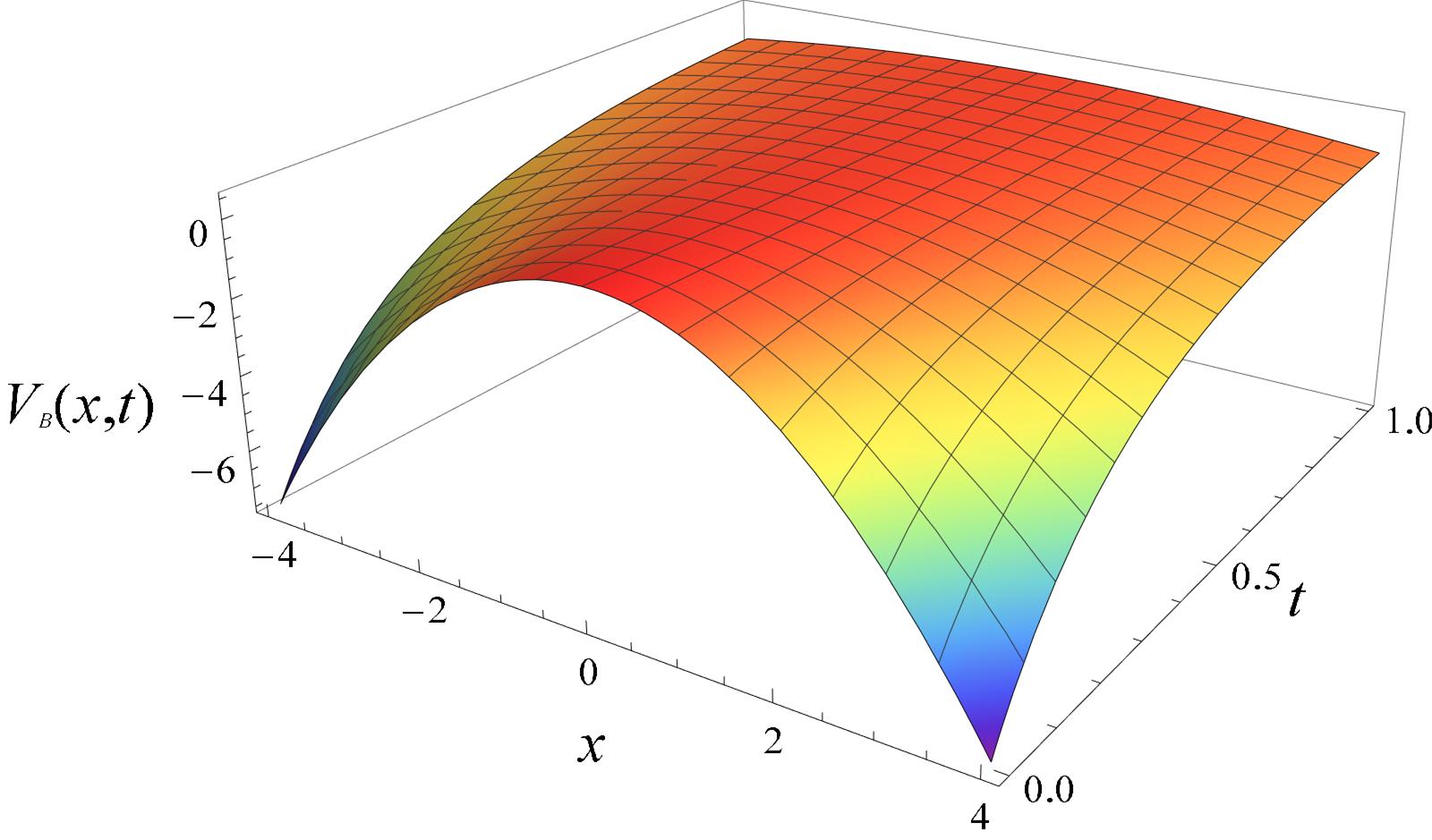}} 
\caption{The Bohm potential, Eq. (\ref{0310}).}
\label{fig3}
\end{figure}

The solution of the Ermakov equation for such frequency may be found to be
\begin{equation}
\rho(t)=\sqrt{a+2t}\sqrt{1+\frac{1}{4}\ln^2(a+2t)}    
\end{equation}
and the value for the parameter $a=1$ is set by the condition $\rho(0)=1$, yielding the time dependent frequency $\Omega(t)=(1+2t)^{-1}$; so,
\begin{equation}
\rho(t)=\sqrt{1+2t}\sqrt{1+\frac{1}{4}\ln^2(1+2t)}.
\end{equation}
Therefore, the Bohm potential results to be
\begin{equation}\label{0310}
V_B(x,t)=-\frac{x^2}{2(1+2t)^2\left[1+\frac{1}{4}\ln^2(1+2t)\right]^2}+\frac{1}{2(1+2t)\left[1+\frac{1}{4}\ln^2(1+2t)\right]};
\end{equation}
hence, there is a kind of phase transition in the system as for $b\rightarrow 2$ the Bohm potential (\ref{0270}) is zero, while for $b=2$ it has finite values (\ref{0310}), as can be seen in Fig.~\ref{fig3}. 

\section{Data Availability}
The data that supports the findings of this study are available within the article

\section{Conclusion}
We have shown that for a phase of the form
\begin{equation}
    S(x,t)=\frac{x^2}{2}\dot{\nu}(t)+\mu(t)
\end{equation}
in the Madelung-Bohm formalism, a Bohm potential that corresponds to the time dependent harmonic oscillator is obtained. The condition for this is that the quantity $e^{\nu(t)}$ obeys the Ermakov equation. A solution via an Ansatz was given in terms of operators that form a closed algebra and in the example we give, namely a rational function of time, it was shown that, because the Ermakov equation presents to different solutions, the Bohm potential also presents two different solutions, one for  $b< 2$ and a different one for $b=2$.

\end{document}